# Analysis of a Triple-cavity Photonic Molecule Based on Coupled Mode Theory

Chao Yang[1], Yong Hu[1], Xiaoshun Jiang[1*], and Min Xiao[1,2]

[1]National Laboratory of Solid State Microstructures, College of Engineering and Applied Sciences, and School of Physics, Nanjing University, Nanjing 210093, China.

[2]Department of Physics, University of Arkansas, Fayetteville, Arkansas 72701, USA.

*Correspondence to: jxs@nju.edu.cn.

PACS number(s): 42.60.Da, 42.82.Et, 42.25.Hz, 42.50.Hz

**Abstract**

In this paper, we analyze a chain-linked triple-cavity photonic molecule (TCPM) with controllable coupling strengths between the cavities on their spectral properties and field (energy) distributions by solving eigenvalues and eigenvectors of the Hamiltonian matrix based on coupled mode theory. Phase transition is extended from double-cavity photonic molecules (DCPMs) to TCPMs, and evolutions of the supermode frequencies and linewidths are analyzed, which have synchronous relations with the degree of coherence between adjacent optical microcavities and energy distributions in the three cavities, respectively. We develop a superposition picture for the three supermodes of the TCPM, as interferences between supermodes of sub-DCPMs. In particular, we demonstrate the abnormal properties of the central supermode in TCPMs, such as dark state in middle cavity and phase shift when energy flowing between side cavities, which are promising in information processing and remote control of energy. General properties of TCPMs are summarized and limitation on linewidths are given. Finally, we make an interesting analog to intracavity electromagnetically induced transparency in multi-level atomic systems using the flexible TCPM platform under appropriate conditions.

# I. INTRODUCTION

Due to the confinement and resonant enhancement of light, optical microcavities [1] have brought great improvements in many aspects of modern photonics. Systems with a large number of interacting optical microcavities, called coupled resonator optical waveguides (CROWs) [2], have aroused great blossom in integrated photonics [3] because of their useful collective properties. Meanwhile, structures with a small number of coupled microcavities, owning to the similarities between the eigenvalues of their supermodes and energy levels in multi-atom molecules, have often been referred to as photonic molecules (PMs) [4,5]. As a simplest example, the double-cavity PM (DCPM) has been extensively studied in various experimental platforms, including coupled semiconductor Fabry-Perot cavities [4], microdisks [6], microspheres [7], and microtoroids [8], covering many promising phenomena such as analog to electromagnetically induced transparency (EIT) [9-11], phonon laser [8], parity-time (PT) symmetry [12,13], generation of exceptional point (EP) [14,15] and so on. Coupled mode theory (CMT) [16,17] is commonly used to theoretically investigate properties of DCPM systems, especially the PT-symmetric [12,18] and EP [14,15,19] behaviors. Although the triple-cavity PMs (TCPMs) have also been studied both theoretically and experimentally in recent years, they were mainly carried out in fixed coupled cavity systems like photonic crystal defect cavities [20], microrings [21,22] and microdisks [6,23]. Recently, by choosing certain parameters in a triple-cavity coupling system, universal sign control is shown to be useful in investigating topological effects [24] while the coupling strengths between the cavities are still fixed. Very recently, we have constructed a fully controllable TCPM platform with two types of coupling configurations (see Fig.1 (c)) [25]. Interesting transmission spectra due to supermode evolutions and dark states have been experimentally demonstrated, and certain spectral properties, including the relative phases between the cavities of the TCPMs, have been explained [6,20,23] using CMT [16,17]. However, large amount of physics behind the supermode evolutions are still not explored in such tunable TCPM system, like whether there is a phase transition (or the so called exceptional point [26]) like in DCPM, how degree of coherence [27] between the fields in different cavities changes with different coupling strengths and how energy flows between the cavities. Typically, finite element method (FEM) [28] and finite difference time domain (FDTD) [29] technique are used to obtain the amplitude and phase information in each cavity [6,20,23,30], which are precise but time-consuming tools. Also, these methods will encounter great difficulty when more coupled cavities are involved. To simplify the theoretical analysis and prepare for extension to more general cases, we recall an alternative technique based on complex Hamiltonian matrix [26,31], and employ eigenvalues and eigenvectors of the Hamiltonian for the triple-cavity coupled system to fully illustrate the supermode frequencies with linewidths and the field distributions inside all three cavities. The physical behaviors around EPs in the TCPM are analyzed and supermode field distributions are well explained in the interference picture, which helps us to better understand this TCPM system, especially with varied coupling strengths.

This paper is organized as following: In Section II, we first construct the Hamiltonian of an N-cavity PM and give its general properties. In Section III, as a reference we reexamine the bonding/antibonding modes of a DCPM first under the identical cavity case (see Fig. 1 (a)-(b)) and then for the unequal loss case. Section IV describes the Hamiltonian approach for the TCPM system. The case of identical cavities (see Fig.1 (c)-(d)) is discussed first to see the fundamental properties of the TCPM's supermodes with variable coupling strengths. Then, we consider cases with certain additional loss in the cavity #2 and with unequal losses in all three cavities, respectively. In Section V, we describe the interesting analog between the intracavity EIT spectrum and the spectrum of a TCPM under appropriate conditions. Section VI serves as a conclusion.

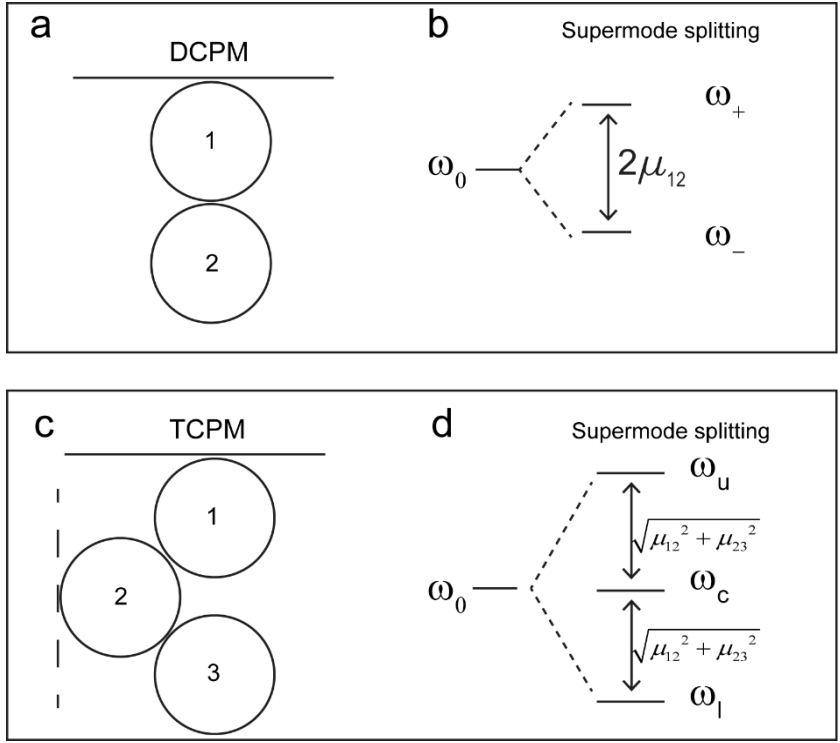

FIG.1 Schematic diagrams of typical DCPM and TCPM and their supermode splittings. In (a) and (c) the circles represent the numbered WGM (whispering gallery mode) cavities and the lines represent waveguides side-coupled for signal input/output. The dashed line in (c) indicates that for a TCPM, there are two types of waveguide coupling schemes, i.e. waveguide coupled with the cavity #1 (Type I) or with the middle cavity #2 (Type II). (b) and (d) indicate the supermode splitting(s) related to the coupling strength(s) in the identical cavity models for DCPM and TCPM, respectively. $\mu_{mn}$ is the coupling strength between cavity #m and cavity #n. The losses of the cavities are neglected here.

## II. THE HAMILTONIAN OF A GENERAL PHOTONIC MOLECULE

First, let's look at the time evolution equations of a typical DCPM [17]. The DCPM is consisted of two WGM microcavities and one waveguide coupled to cavity #1 as the signal input/output ports (Fig.1 (a) [8]). The equations governing the time evolutions of the two cavity modes [17] can be written as

$$\begin{cases} \dfrac{d}{dt}a_1 = (i\Delta\omega_1 - \gamma_1)a_1 - i\mu_{12}a_2 + \sqrt{\kappa_1}s_{in} \\ \dfrac{d}{dt}a_2 = -i\mu_{12}^*a_1 + (i\Delta\omega_2 - \gamma_2)a_2 \end{cases}. \qquad (1)$$

Here we denote the field amplitude, the frequency detuning and the total loss of the nth cavity as $a_n$, $\Delta\omega_n$ and $\gamma_n$, respectively, where $\Delta\omega_n = \omega_n - \omega$ with n=1,2. The coupling strengths between the mth and nth cavities, and between the nth cavity and the waveguide, are defined as $\mu_{mn}$ and $\kappa_n$, respectively. $|s_{in}|^2$ is the input power. $i^2 = -1$. The solutions of these equations give the field amplitudes. We can rewrite equation (1) into a matrix form $i\frac{d}{dt}\vec{A_2} = \vec{M_2}\vec{A_2} - \vec{B_2}$, where the subscript marks the dimension of the matrix. $\vec{A_2} = [a_1 \ a_2]^T$ and $\vec{B_2} = [-i\sqrt{\kappa_1}s_{in} \ 0]^T$ are the intracavity field vector and the input field vector, respectively. The coefficient matrix for the DCPM now becomes

$$\vec{M_2} = \begin{pmatrix} \tilde{\beta}_1 & \mu_{12} \\ \mu_{12}^* & \tilde{\beta}_2 \end{pmatrix} - \omega \vec{I_2} = \vec{H_2} - \omega \vec{I_2}, \tag{2}$$

where $\vec{H_2}$ is the Hamiltonian of the double-cavity coupling system [31], $\tilde{\beta}_n = \omega_n - i\gamma_n$ represents the initial resonant frequency and total optical loss of the nth cavity, and $\vec{I_2}$ marks the two-dimensional unit matrix.

In fact, the Hamiltonian of an N-cavity PM can be easily generalized to be $\vec{H_N} = \vec{M_N} + \omega \vec{I_N}$, the subscript N denotes the dimension of the matrices and vectors. The diagonal elements are $H_{nn} = \tilde{\beta}_n$, indicating each cavity's initial condition, and the off-diagonal element $H_{mn} = \mu_{mn}$ indicates the interaction between the mth and nth cavities, which is Hermitian ($\mu_{mn}^* = \mu_{mn}$) if the coupling is assumed to be lossless [16]. Next, the supermode frequencies with linewidths, as well as the field amplitudes with phase information, can be easily obtained by solving for the eigenvalues and constructing linear superpositions of eigenvectors as determined by the input vector [31].

## III.  THE SUPERMODES OF A DCPM

As aforementioned, we can easily solve the eigenvalues and eigenvectors of the DCPM's Hamiltonian (Eq. (2)) in general to have:

$$\begin{cases} \tilde{\omega}_{\pm} = \dfrac{\tilde{\beta}_1 + \tilde{\beta}_2}{2} \pm \sqrt{(\dfrac{\tilde{\beta}_1 - \tilde{\beta}_2}{2})^2 + \mu_{12}^{\;2}} \\ \vec{A}_{\pm} = \left[ \dfrac{\tilde{\beta}_1 - \tilde{\beta}_2 \pm \sqrt{(\tilde{\beta}_1 - \tilde{\beta}_2)^2 + 4\mu_{12}^{\;2}}}{2\sqrt{2}\mu_{12}} \quad \dfrac{1}{\sqrt{2}} \right]^T \end{cases} . \qquad (3)$$

The subscript " $\pm$ " indicates the bonding/antibonding modes, respectively. The eigenvalues $\tilde{\omega}_{\pm}$ are complex, and their real and imaginary parts correspond to the frequencies and linewidths of supermodes, respectively. The eigenvectors $\vec{A}_{\pm}$ are field vectors of the bonding/antibonding modes, through which we can figure out the amplitude and phase information of fields in the cavities. We note that the eigenvectors are naturally orthogonal.

**3.1 DCPM with two identical cavities**

First, we consider the case when the two cavities in Fig. 1 (a) are identical, i.e., $\tilde{\beta}_1 = \tilde{\beta}_2 = \tilde{\beta}_0 = \omega_0 - i\gamma_0$, so that the eigenvalues and the eigenvectors are reduced down to [26]:

$$\begin{cases} \tilde{\omega}_{\pm} = \omega_0 \pm \mu_{12} - i\gamma_0 \\ \vec{A}_{\pm} = \begin{bmatrix} \pm 1 & 1 \end{bmatrix}^T / \sqrt{2} \end{cases} . \qquad (4)$$

In this case, the supermode splitting is proportional to the coupling strength $\mu_{12}$ (as shown in Fig.1 (b)) and for both supermodes, the fields evenly distribute in the two cavities while the phase difference (PD) is kept to be either 0 (for the bonding mode) or $\pi$ (for the antibonding mode) for any coupling strength. As a result, the coherence of the fields [27] in the two cavities maintains to be constructive (destructive) in the bonding (antibonding) mode. The imaginary parts of the two supermodes remain to be $\gamma_0$ in this case, thus the spectral linewidths of the bonding/antibonding modes are equal and constant. Although this case is quite ideal, it can well explain the basic properties of a general DCPM system in the strong coupling limit, such as mode splitting and field distribution/coherence of the supermodes.

**3.2 DCPM with unequal cavity losses**

Next, we consider the case with cavities having an equal resonant frequency but different losses, i.e., $\omega_1 = \omega_2 = \omega_0$ and $\gamma_1 \neq \gamma_2$. By setting $\gamma_d = |\gamma_1 - \gamma_2|/2$ and $\gamma_{ave} = (\gamma_1 + \gamma_2)/2$, Eq.(3) is rewritten as

$$\mu_{12} < \gamma_d : \begin{cases} \tilde{\omega}_\pm = \omega_0 + i(\gamma_{ave} \pm \sqrt{\gamma_d^2 - \mu_{12}^2}) \\ \vec{A}_\pm = [i(\sin\varphi \mp \cos\varphi) \quad 1]^T / \sqrt{2} \end{cases}$$
$$\mu_{12} \geq \gamma_d : \begin{cases} \tilde{\omega}_\pm = \omega_0 \pm \sqrt{\mu_{12}^2 - \gamma_d^2} + i\gamma_{ave} \\ \vec{A}_\pm = [i\sin\varphi \mp \cos\varphi \quad 1]^T / \sqrt{2} \end{cases}. \quad (5)$$

Here, we introduce a characteristic angle $\varphi = \arctan(\gamma_d / \sqrt{|-\gamma_d^2 + \mu_{12}^2|})$ to better describe the phase transition. It's clear that a transition point emerges when $\mu_{12} = \gamma_d$ and $\varphi = \pi/2$ (see Fig.2). The characteristic angle is equal to the phase difference between the two cavities when $\mu_{12} \geq \gamma_d$ (see Fig.2 (d)).

Figure 2 shows how the coupling strength affects $\tilde{\omega}_\pm$ and field distributions in different cavities. In Fig. 2 (a), the real (imaginary) parts of $\tilde{\omega}_\pm$ are plotted as the red and blue thick (dashed) lines, respectively. Similar behaviors have been reported previously in papers related to exceptional points [18,19]. Here, with the eigenvectors of Hamiltonian matrix, we calculate field vector evolutions and plot the degree of coherence (DOC) [27] between the two cavities as defined by $\text{DOC}_\pm = (a_{1\pm} + a_{2\pm})^2 - 1$ (red dotted lines. It equals to zero if completely incoherent, and $\pm 1$ if fully coherent) in Fig.2 (c) and the field distribution in the first cavity of DCPM as represented by the parameter $\eta_\pm = a_{1\pm}^2 / (a_{1\pm}^2 + a_{2\pm}^2)$ (blue point lines) in Fig.2 (d). The spectral properties of a DCPM (i.e., splitting of the supermode frequencies and evolutions of linewidths) are synchronously connected with field information (i.e., DOC and field distributions), which indicates:

*Note (i)*: the DOC between fields in the two cavities, as well as the separation distance between the cavities, contribute to the supermode splitting, because the interference in the coupling region affects the effective refractive indices of the supermodes [1].

*Note (ii)*: the field distributions in the cavities influence the linewidths of the supermodes, since the field energy in the cavity with a bigger loss decays faster which corresponds to a larger linewidth.

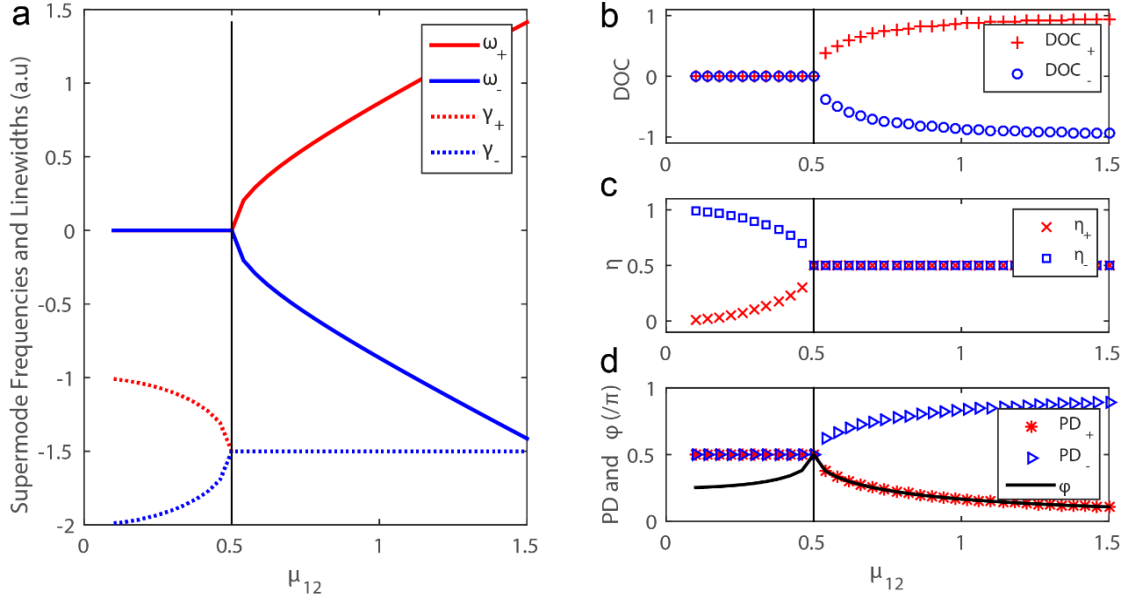

FIG.2 (Color online) The spectral properties and field information of bonding (red plots) and antibonding (blue plots) modes for a DCPM with unequal cavity losses, as functions of $\mu_{12}$: (a) the supermode frequencies (lines) and their linewidths (dashed lines), (b) degrees of coherence (DOC, the crosses and circles) and (c) field distribution parameter $\eta$ (× and squares), (d) phase difference (PD, the asterisk and triangles) and characteristic angle $\varphi$ (the black line). Note that field information for the two modes are represented by different markers in case they overlap and one branch becomes indiscernible. The parameters are: $\gamma_1 = 2$, $\gamma_2 = 1$, $\omega_0 = 0$. The black vertical lines mark the EP points where $\mu_{12} = \gamma_d$. (a), (b) and (c) reveal that the behaviors of $\omega_\pm$ and $\gamma_\pm$ synchronize with $DOC_\pm$ and $\eta_\pm$, respectively. (d) shows that $\varphi$ coincides with $PD_+$ after phase transition.

To clearly show the field change in the DCPM around its EP [26], we calculate the phase difference (PD) between the field in cavity #1 and field in cavity #2 ( by setting the phase of field in cavity #2 as a reference, the phase difference equals to the phase of field in cavity #1) and the characteristic angle $\varphi$, as displayed in Fig.2 (d). When there is no coupling (i.e. $\mu_{12} = 0$), $\varphi = \pi/4$, so the phases of $a_{1\pm}$ are both $\pi/2$ and $DOC_\pm = 0$, $\gamma_+ = \gamma_2$, $\gamma_- = \gamma_1$, the structure acts as two isolated cavities. As $\mu_{12}$ increases to $\gamma_d$, $\varphi$ grows up to $\pi/2$, so $a_{1\pm}$ transform from $i(\sin\varphi \mp \cos\varphi)$ to $i\sin\varphi \mp \cos\varphi$ (see Eq.(5)) and half of the bonding (antibonding) mode energy in cavity #1 (#2) transfers to cavity #2 (#1), with $\gamma_+ = \gamma_- = \gamma_{ave}$. At $\mu_{12} = \gamma_d$, two important changes occur around the EP [26]:

*Change (i)*: the interaction between the two cavities is so strong that the field energies distributed in the two cavities are equal and keep to be the same afterwards [4]. At the same time, the linewidths of supermodes stop to change.

*Change (ii)*: fields in the two cavities stop being incoherent [5]. The characteristic angle coincides with the phase difference for the symmetric mode between the two cavities (Figs.2 (c) and (d)) and the coupling-induced mode splitting starts to occur.

When $\mu_{12} > \gamma_d$, though the field amplitudes keep unchanged, the phase differences between the two cavities in the bonding and anti-bonding modes ($PD_\pm$) start to deviate from $\pi/2$. With $\mu_{12}$ increasing, $PD_+ = \varphi$ decreases and $PD_- = \pi - \varphi$ increases, as shown in Fig.2 (c). *The constructive and destructive interferences* in the inter-cavity coupling region affect the field amplitudes in the coupling region, especially at the gap where the refractive index is smaller (no medium) and thus the effective refractive index [1] is inevitably modified by $\varphi$. To satisfy resonant condition, the supermode frequencies split accordingly (red thick lines in Fig.2 (a)).

Under the deep strong coupling condition of $\mu_{12} \gg \gamma_d$, $\varphi \to 0$, $PD_+ \to 0$, $PD_- \to \pi$ and $\sqrt{\mu_{12}^2 - \gamma_d^2} \to \mu_{12}$; the supermode frequencies, linewidths, and the field vectors all reduce down to the case of identical cavity model (see Eq. (4)). It's clear that the results for unequal losses will coalesce with the identical cavity case under deep strong coupling condition.

Furthermore, by doping active material into cavity #1, the DCPM becomes a typical parity-time (PT) symmetric system [32] when the gain in the active cavity balances the loss in the passive cavity [12,13], i.e. $\gamma_1 = -\gamma_2 = -\gamma_0$. The EP occurs at the coupling strength value of $\mu_{12}=\gamma_0$ [19], and the linewidth of the supermode in the strong coupling regime ($\mu_{12} \geq \gamma_0$) becomes zero. In other words, the eigenvalues of the non-Hermitian Hamiltonian reduce down to a real number which reaches the so called unbroken PT-symmetric phase [33]. One can consider that case as an extension of typical DCPMs with $\gamma_d=\gamma_0$ and $\gamma_{ave}=0$ [12,13,18]. Note that the evolutions of the phase relations and energy distributions in the PT-symmetric PMs are the same as the typical DCPMs since $\varphi$ is independent of $\gamma_{ave}$.

## IV. SUPERMODES OF A TCPM

For a TCPM, we denote the three supermodes as *central, upper,* and *lower* modes, respectively. Recently, two types of TCPM have been constructed, as shown in Fig.1 (c), and with variable coupling strengths, supermode evolutions of the TCPMs have been studied and mode splittings in strong coupling regime have been well fitted using CMT equations like the one shown in Fig.1 (d) [25]. In the following, using the Hamiltonian matrix formulism, we analyze the evolutions of the supermode frequencies, linewidths and the field distributions, and reveal the interference relations of the cavity fields and energy flowing within the TCPM (Type I).

Since there is no direct interaction between cavity #1 and cavity #3 for the chain-linked (Type I) TCPM (i.e. $\mu_{13}^*=\mu_{13}=0$), the Hamiltonian matrix of the system (see Section II) can be expressed in the form [16]:

$$\overrightarrow{H_3} = \begin{pmatrix} \tilde{\beta}_1 & \mu_{12} & 0 \\ \mu_{12}^* & \tilde{\beta}_2 & \mu_{23} \\ 0 & \mu_{23}^* & \tilde{\beta}_3 \end{pmatrix} . \quad (6)$$

The cavity field vector becomes $\overrightarrow{A_3} = \begin{bmatrix} a_1 & a_2 & a_3 \end{bmatrix}^T$ and $\overrightarrow{B_3}$ is the input field vector. The eigenvalues are the roots of the characteristic equation:

$$\Omega(\omega) = (\tilde{\beta}_1 - \omega)(\tilde{\beta}_2 - \omega)(\tilde{\beta}_3 - \omega) - (\tilde{\beta}_3 - \omega)\mu_{12}^2 - (\tilde{\beta}_1 - \omega)\mu_{23}^2 = 0 . \quad (7)$$

Although the expressions of the roots are complex in general, one can still see some important properties in (7). One property to notice is:

$$\tilde{\omega}_u + \tilde{\omega}_c + \tilde{\omega}_l = \tilde{\beta}_1 + \tilde{\beta}_2 + \tilde{\beta}_3, \qquad (8)$$

where $\tilde{\omega}_{u,c,l}$ are complex numbers, denoting the frequencies (real parts) and linewidths (imaginary parts) of the three supermodes, respectively. Equation (8) indicates that the sum of supermode frequencies and the sum of their linewidths equal to the sum of the uncoupled cavities' initial resonant frequencies and losses, respectively. To better see the essential behaviors, we first analyze properties of the TCPM supermodes in the ideal case.

**4.1. TCPM with three identical cavities**

Let's first consider the ideal case of three identical cavities (i.e. $\tilde{\beta}_1 = \tilde{\beta}_2 = \tilde{\beta}_3 = \tilde{\beta}_0 = \omega_0 - i\gamma_0$). Equation (7) then becomes $(\tilde{\beta}_0 - \omega)[(\tilde{\beta}_0 - \omega)^2 - \mu_{12}^2 - \mu_{23}^2] = 0$ and the eigenvalues are easily solved to be $\tilde{\omega}_c = \omega_0 - i\gamma_0$ and $\tilde{\omega}_{u,l} = \omega_0 \pm \sqrt{\mu_{12}^2 + \mu_{23}^2} - i\gamma_0$. Similar to DCPM (see Eq. (4)), the linewidths of the modes are all $\gamma_0$. The upper/lower mode frequencies, splitting by $\pm\sqrt{\mu_{12}^2 + \mu_{23}^2}$, respectively, are affected only by the two coupling strengths, while the central mode maintains at the initial resonant frequency.

The eigenvectors (without normalization), which directly show the interference effects in the TCPM, are given by

$$\vec{A}_c = [-\mu_{23} \quad 0 \quad \mu_{12}]^T, \quad \vec{A}_{u,l} = [\mu_{12} \quad \pm\sqrt{\mu_{12}^2 + \mu_{23}^2} \quad \mu_{23}]^T. \qquad (9)$$

As Fig.3 shows, the upper (lower) mode is resulted from the constructive interference (in the middle cavity) of two sub-DCPMs with symmetric (anti-symmetric) mode, having the same phase (zero) for the fields in two side cavities. A previous work has predicted that the central mode is a superposition between one symmetric mode in one sub-DCPM and one anti-symmetric mode in another sub-DCPM, and the two DCPMs share the middle cavity [34]. However, the current solution doesn't support that prediction: the fields in cavities #1 and #3 have opposite sign while in the cavity #2 is zero. In fact, as Fig.3 shows, for the central mode, $a_1$ ($a_3$) grows with $-\mu_{23}$ ($\mu_{12}$) while $a_2=0$, which has been confirmed by the experimental results in the controllable TCPM system [25]. This seems to be abnormal because a remote control on field amplitudes of side cavities is realized: when we strengthen $\mu_{23}$ by moving cavity #3 closer to cavity #2, the field amplitude (energy) in cavity #1 increases. This result can be easily understood if we take all three eigenvectors into consideration and regard them as a new set of basis vectors for the TCPM system, which are orthogonal to each other. In Fig.3 we plot, in a blue background, the superposition picture of central mode to indicate that it's quite different from the upper and lower modes.

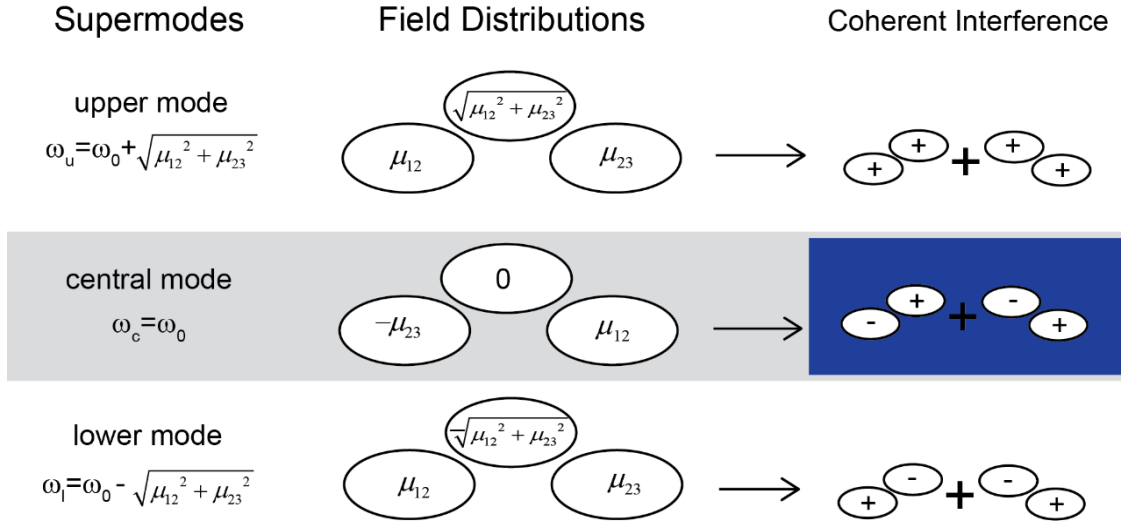

FIG.3 The superpositions of sub-DCPM modes for the upper, central, and lower modes in TCPM. The upper (lower) mode is a result of constructive interference between two symmetric (asymmetric) modes of the sub-DCPMs, while the central mode reveals a dark state in cavity #2 and an exchanged dependence on the coupling strengths: $a_1$ ($a_3$) grows with $-\mu_{23}$ ($\mu_{12}$).

To see the field evolutions better, we normalize Eq. (9) as following:

$$\begin{cases} \vec{A}_u = [\cos\theta \quad 1 \quad \sin\theta]^T / \sqrt{2} \\ \vec{A}_c = [-\sin\theta \quad 0 \quad \cos\theta]^T \\ \vec{A}_l = [\cos\theta \quad -1 \quad \sin\theta]^T / \sqrt{2} \end{cases}, \quad (10)$$

where $\theta = \arctan(\mu_{23}/\mu_{12})$. Different from the DCPM case with identical cavities where the field amplitudes do not change, here $a_1$ and $a_3$ change with $\theta$, while $a_2$ keeps to be constant ($a_{2,c}=0$ and $a_{2,u/l}=\pm 1/\sqrt{2}$) due to destructive and constructive interferences. Another important difference between TCPM and DCPM is that field distributions within the TCPM can be modulated by $\theta$ while keeping the mode splitting fixed by carefully controlling the coupling strengths. What's more, as aforementioned, the eigenvectors in TCPM are also naturally orthogonal.

Figures 4 (a)-(c) show the normalized field amplitudes $a_n$ and (d)-(e) the energy distributions $\eta_n = a_n^2 / \sum_n a_n^2 = |a_n|^2$ in the nth (n=1,2,3) cavities of the supermodes as a function of $\theta$. The red, green and blue lines illustrate the field amplitudes or energies in the cavities #1, #2 and #3, respectively. When $\theta$=0 ($\mu_{23} \ll \mu_{12}$), $a_{1,(u,l)}=1/\sqrt{2}$, $a_{2,(u,l)}=\pm 1/\sqrt{2}$, $a_{3,c}=1$, $a_{3,(u,l)}=a_{(1,2),c}=0$, the upper and lower mode energies distribute evenly in cavities #1 and #2 ($\eta_{1,(u,l)}=\eta_{2,(u,l)}=1/2$), and the central mode energy totally localizes in cavity #3 ($\eta_{3,c}=1$). That is, the DCPM, composed of cavities #1 and #2, has little interaction with cavity #3. With $\theta$ increasing, field (energy) in cavity #2 stays the same, while energy transfers between cavity #1 and cavity #3: $\eta_{1,(u,l)}$ decreases while $\eta_{3,(u,l)}$ increases without changing field amplitude sign, i.e., energy flows from cavity #1 to cavity #3 without changing field phases. Meanwhile, $\eta_{3,c}$ decreases and $\eta_{1,c}$ increases and, since $a_{3,c}$ and $a_{1,c}$ have different signs, energy flows from cavity #3 to cavity #1 with a $\pi$ phase shift for the field after flowing. When $\theta=\pi/4$ ($\mu_{23}=\mu_{12}$), energy distributions of the structure in all the supermodes are symmetric ($\eta_{1,(u,c,l)}=\eta_{3,(u,c,l)}$), and field distributions in the upper and lower modes become symmetric ($a_{2,(u,l)}=\pm 1/\sqrt{2}$, $a_{1,(u,l)}=a_{3,(u,l)}=1/2$) while in the central mode anti-symmetric ($a_{1,c}=-1/\sqrt{2}, a_{3,c}=1/\sqrt{2}$), which are supported by numerical calculations and near-field mapping with fixed coupling structures [20,23,30]. When $\theta=\pi/2$ ($\mu_{23} \gg \mu_{12}$), energies in the upper and lower modes distribute equally in cavities #2 and #3 ($\eta_{2,(u,l)} = \eta_{3,(u,l)} = 1/2$) while in the central mode totally in cavity #1 ($\eta_{1,c} = 1$), i.e., cavity #2 and cavity #3 form another DCPM and cavity #1 becomes isolated. We note the range of parameter $\theta \in (0, \pi/2)$. From Fig.4 (b), one can see that a $\pi$ phase shift of field (amplitude sign changes from positive to negative) only happens in the central mode during energy flowing from cavity #3 to cavity #1, while no phase shifts in the upper/lower modes.

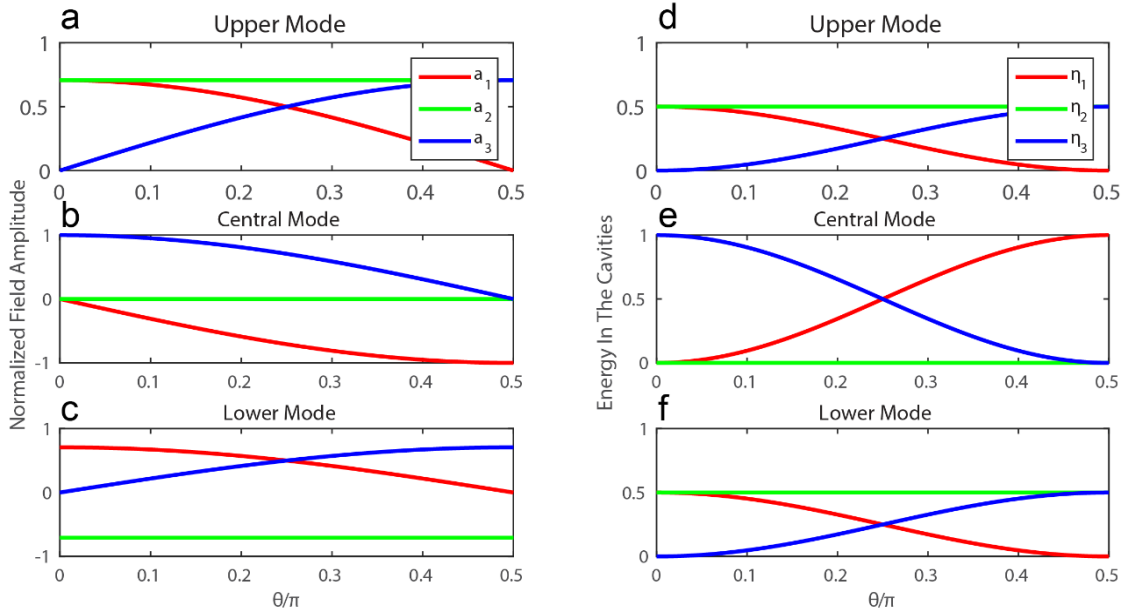

FIG.4 (Color online) (a)-(c) The normalized field amplitudes and (d)-(f) the energy distributions in the three identical cavities in the upper, central and lower modes, respectively, as a function of $\theta$. The red, green and blue lines illustrate the field amplitudes and energy distributions in the cavities #1, #2 and #3, respectively. With $\theta$ increases, field (energy) flows between cavity #1 and #3 with cavity #2 undisturbed. A sign change between $a_1$ and $a_3$ in the central mode as shown in panel (b), indicates a $\pi$ phase shift of field after energy flowing.

### 4.2 TCPM with a certain loss in the middle cavity

Next, we consider the case with a certain loss $\Delta\gamma_2$ in the middle cavity of a Type I TCPM [25] (for Type II TCPM, the central mode can't be excited), i.e., $\tilde{\beta}_1 = \tilde{\beta}_2 - i\Delta\gamma_2 = \tilde{\beta}_3 = \tilde{\beta}_0$. The eigenvalues and eigenvectors are given as following:

$$\tilde{\omega}_c = \tilde{\beta}_0, \quad \vec{A}_c = \begin{bmatrix} -\sin\theta & 0 & \cos\theta \end{bmatrix}^T \qquad (11)$$

and

$$\sqrt{\mu_{12}^2 + \mu_{23}^2} < \gamma_d' : \begin{cases} \tilde{\omega}_{u,l} = \omega_0 - i\gamma_{ave}' \pm i\sqrt{\gamma_d'^2 - (\mu_{12}^2 + \mu_{23}^2)} \\ \vec{A}_{u,l} = [\cos\theta \quad \pm i(\sin\varphi' \mp \cos\varphi') \quad \sin\theta]^T / \sqrt{2} \end{cases} \quad (12.a)$$

$$\sqrt{\mu_{12}^2 + \mu_{23}^2} \geq \gamma_d' : \begin{cases} \tilde{\omega}_{u,l} = \omega_0 - i\gamma_{ave}' \pm \sqrt{\mu_{12}^2 + \mu_{23}^2 - \gamma_d'^2} \\ \vec{A}_{u,l} = [\cos\theta \quad \pm (i\sin\varphi \mp \cos\varphi') \quad \sin\theta]^T / \sqrt{2} \end{cases}, \quad (12.b)$$

where $\gamma_d' = \Delta\gamma_2/2$, $\gamma_{ave}' = \gamma_0 + \Delta\gamma_2/2$ and $\varphi' = \arctan(|\gamma_d'/\sqrt{\mu_{12}^2 + \mu_{23}^2 - \gamma_d'^2}|)$. Figure 5 shows the evolutions of the complex eigenvalues when $\mu_{23}$ increases under the condition of $\mu_{12} < \gamma_d'$. The real parts (supermode frequencies, see Fig.5 (a)) of the upper/lower modes $\omega_{u,l}$ split and the imaginary parts (supermode linewidths, see Fig.5 (b)) $\gamma_{u,l}$ coalesce at a certain point $\mu_{23} = \sqrt{\gamma_d'^2 - \mu_{12}^2}$, while the central mode keeps unchanged in both frequency and linewidth. The parameters are: $\gamma_1 = \gamma_3 = 1$, $\gamma_2 = 3$, $\mu_{12} = 1/2$, and, at the transition point, $\mu_{23} = \sqrt{3}/2$.

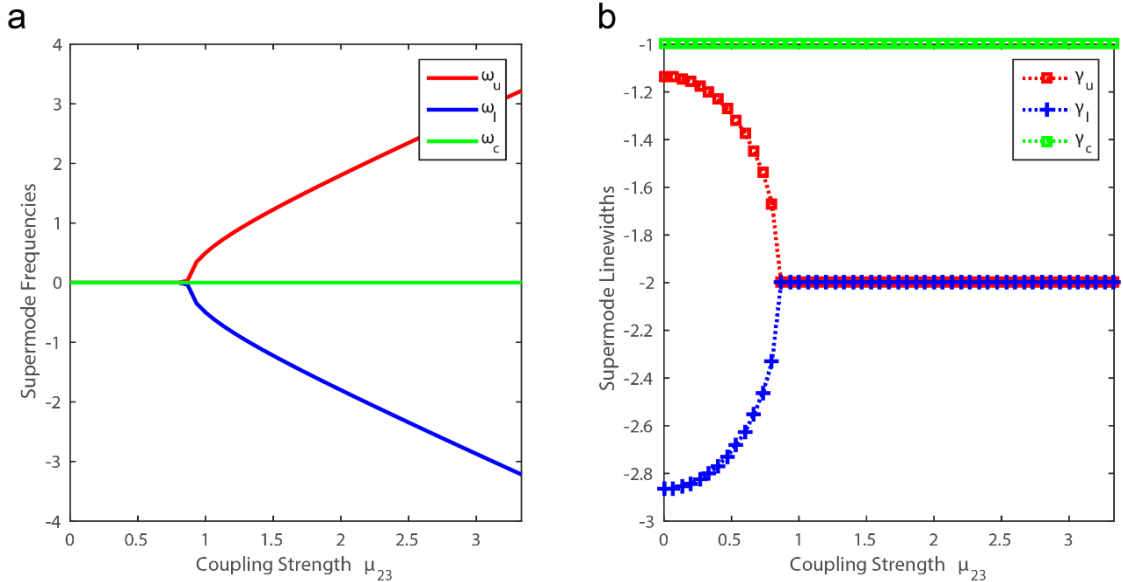

FIG.5 (Color online) Evolutions of the complex eigenvalues of a TCPM Hamiltonian with $\mu_{23}$ increases: (a) the real parts $\omega_{u,l,c}$ (supermode frequencies) and (b) imaginary parts $\gamma_{u,l,c}$ (supermode linewidths). The red, blue and green lines and markers denote the upper, lower and central modes of the TCPM with parameters: $\gamma_1 = \gamma_3 = 1$, $\gamma_2 = 3$ and $\mu_{12} = 1/2$. A phase transition (or EP) occurs at $\mu_{23} = \sqrt{3}/2$, which satisfies $\sqrt{\mu_{12}^2 + \mu_{23}^2} = \gamma_d'$.

Recalling *notes (i) and (ii) and changes (i) and (ii)* in Section 3.2 for DCPM, we can naturally find the relevant field information, based on the eigenvectors in Eq.(11) and Eq.(12), to understand the supermode evolutions in Fig.5: the degree of coherence [27] between adjacent cavities $DOC_{12}$ and $DOC_{23}$ (generally defined by $DOC_{mn} = (a_m + a_n)^2 / (a_m^2 + a_n^2) - 1$) and the energy distribution in the nth cavity ($\eta_n$). We plot $DOC_{12}$ and $DOC_{23}$ in Figs.6 (a)-(c) and $\eta_1$, $\eta_2$, $\eta_3$ in Figs.6 (d)-(f), respectively, as a function of $\mu_{23}$. It's quite clear that the physical picture of interference in the coupling region (see *note (i)* in Section 3.2) and linewidth function of energy distribution $\gamma_s(\eta_n) = \sum_n \gamma_n * f(\eta_n)$ ($f(\eta_n) \approx \eta_n$, regardless of the coupling region between adjacent WGM cavities) can also be applicable to this TCPM case.

Figures 6 (a) and (c) show that $DOC_{12}$ and $DOC_{23}$ of the upper (lower) mode change from zero to positive (negative) values at the transition point, and with $\mu_{23}$ increasing, the absolute value of $DOC_{23}$ increases continuously while the absolute value of $DOC_{12}$ first increases and then decreases back towards zero due to the enlarged difference between $a_1$ and $a_2$. Positive DOC indicates a constructive interference in the coupling region between adjacent cavities while negative one destructive (see Fig.3), and the change of effective refractive index causes the upper/lower mode to split. Since $a_2$ is always zero in the central mode, $DOC_{12} = DOC_{23} = 0$ (see Fig.6 (b)), and $\omega_c$ keeps unchanged.

Figures 6 (d)-(f) show how the energy distribution in each cavity changes with $\mu_{23}$ increase, corresponding to the case in Fig.5. Considering $\gamma_s(\eta_n) \approx \sum_n \gamma_n * \eta_n$ (n=1,2,3), $\eta_{2,c} = 0$, $\gamma_1 = \gamma_3$ when $\sqrt{\mu_{12}^2 + \mu_{23}^2} \geq \gamma_d'$, we can write the linewidths as $\gamma_{u,l}(\eta_2) \approx \gamma_1 + \eta_2 \Delta \gamma_2$ and $\gamma_c = \gamma_1$. Though energy in the central mode flows from cavity #1 to cavity #3, the linewidth of this central mode does not change (see Fig.6 (e)). Now we focus on $\eta_2$ in the upper/lower modes and clearly find that $\eta_2$ decreases (increases) from 1 (0) in the upper (lower) mode, and they coalesce to 1/2 at the transition point.

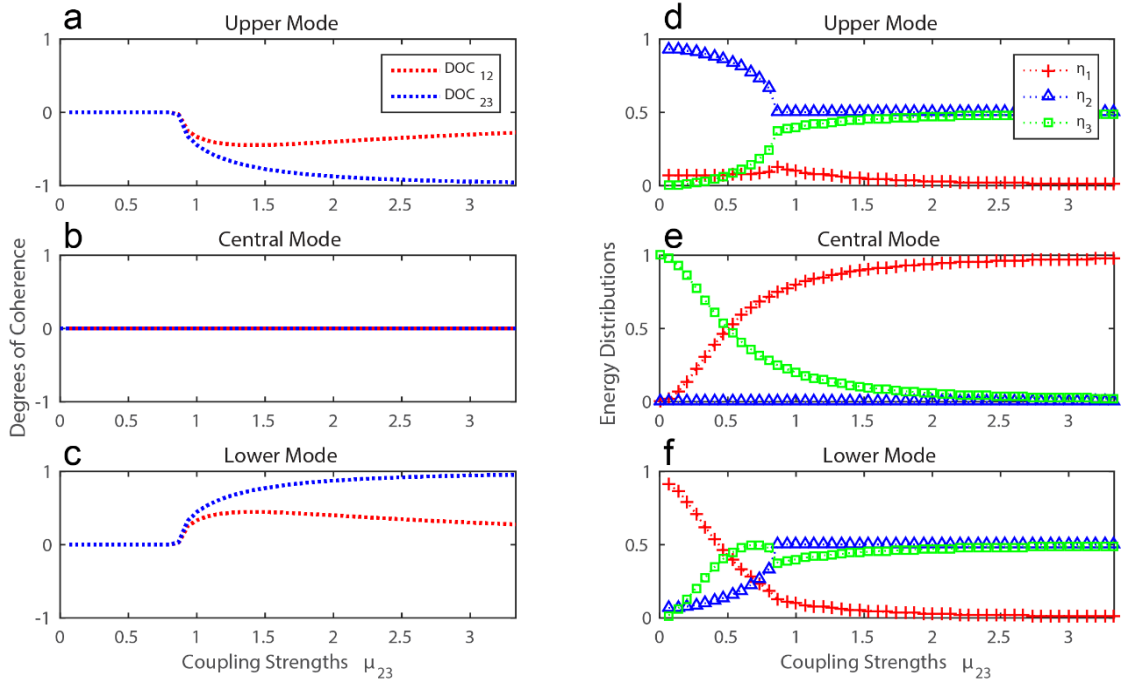

FIG.6 (a)-(c) Degrees of coherence between adjacent cavities and (d)-(f) energy distributions in different cavities for different supermodes of TCPM with middle cavity having a certain loss. The red and blue dashed lines in (a)-(c) denote $DOC_{12}$ and $DOC_{23}$, respectively, and the red crosses, blue triangles and green squares denote the energy distributions in cavities #1, #2 and #3, respectively. The parameters are the same as in Fig.5. Around EP, the DOC and energy distribution reveal the evolutions of field behind the spectral properties (mode splittings and linewidths) of TCPM's supermodes in FIG.5.

By comparing Eq.(11) and Eq.(12.b) with Eq.(10), it is clear that the additional loss causes a phase shift $\varphi'$ in cavity #2 and doesn't affect fields in cavities #1 and #3 when $\sqrt{\mu_{12}^2 + \mu_{23}^2} \geq \gamma_d'$. Thus the central mode just behaves the same way as in Section (4.1). In fact, when very strong coupling condition is satisfied ($\sqrt{\mu_{12}^2 + \mu_{23}^2} >> \Delta\gamma_2$), the phase angle $\varphi' \to 0$, so the case totally reduces to the situation as in Section (4.1). Note that several interesting properties emerge for the central mode: (1) Cavity #2 is in a robust "dark state" [6], i.e., no matter how much loss is introduced in cavity #2, its field won't change. (2) No matter how the coupling strengths change, the central mode frequency and linewidth can't be affected. (3) A $\pi$ phase transition of fields can be realized after transferring field between cavity #1 and cavity #3 when $\theta$ increases from 0 to $\pi/2$. These properties make TCPM a promising platform for quantum information processing and remote information exchange between cavity #1 and cavity #3 if a specially-designed cavity #2 is used to connect them.

**4.3 TCPM with unequal losses**

Finally, we consider the general case with unequal losses $\gamma_1 \neq \gamma_2 \neq \gamma_3$. Under such a condition, the expressions of eigenvalues and eigenvectors become so complex that it wouldn't be useful to write them down here. Nevertheless, based on the physical pictures presented above (Section (3.2): *two notes and two changes*), we can still predict some qualitative tendencies and limitations for the evolutions of TCPM supermodes. (1) In the weak coupling limit (i.e. the coupling strengths between adjacent cavities are small or comparable to the losses of the cavities [31]), there should also be a transition point, where the energy transferring becomes sufficient to change degree of coherence between fields in adjacent cavities and therefore supermode splitting occurs. (2) In the strong coupling limit (i.e. the coupling strengths are much larger than the losses of the cavities [31]), the supermode splittings should be close to $\sqrt{\mu_{12}^2 + \mu_{23}^2}$. Considering linewidth functions of the supermodes and neglecting the contribution of energy in the overlap region of the two cavity fields to linewidths of the supermodes, the supermode linewidths are given by $\gamma_{u,l} \approx (\gamma_1 \cos^2\theta + \gamma_3 \sin^2\theta + \gamma_2)/2$ and $\gamma_c \approx \gamma_1 \cos^2\theta + \gamma_3 \sin^2\theta$, respectively. This implies that both linewidth narrowing and broadening for the central mode can happen if the coupling strengths are properly controlled. (3) Equation (8) is always satisfied for the TCPM system.

Furthermore, since the essence of CMT is for treating coupling of multiple resonances, the above theoretical calculations and conclusions can also be applicable to various three-mode coupled mechanical, atomic and quantum mechanical systems [26,31]. In the same time, this TCPM platform can exhibit interesting behaviors in analog to other resonance-coupled systems, for example, double EIT [35,36] (as studied in four-level atomic systems) and intracavity EIT [37,38] (as observed in systems with three-level atoms inside an optical cavity).

## V. ANALOG TO THE INTRACAVITY EIT PHENOMENON

In Refs. [25,39], analog to double EIT phenomenon has been realized in the coupled triple-cavity systems. In this section, we make an analog between a TCPM with $\gamma_1 \neq \gamma_2 \neq \gamma_3$ and the intracavity EIT system [37,38]. A typical intracavity EIT system is consisted of a three-level atomic (EIT) medium coupled with an optical cavity [37,38], which is also a three-mode coupling system. The spectrum of typical three-peak "dark- and bright-sate polaritons" [40,41] appears in the coupled intracavity three-level atomic system and results in a central-peak-narrowing phenomenon. The much narrowed central peak is caused by enhanced normal dispersion due to EIT [37,38]. This is similar to the dark state of the central mode in TCPM, where the central mode linewidth can be much narrower than the linewidth of cavity #1 if $\gamma_3 < \gamma_1$ [38,40]. Looking back at Section IV, one can see that there are some obvious advantages in using the TCPM platform over atomic systems to realize linewidth narrowing and mode splitting structures, such as no need for a strong pump (coupling) light, controllability on many degrees of freedom, and more importantly being able to integrate on chip for future practical device applications.

Figure 7 shows theoretical plots of field energy distributions ($|a_i|^2$) within a TCPM and the transmission spectrum from a side-coupled waveguide with similar parameters as in Ref [40] for intracavity EIT, i.e. $2\gamma_1$=14 MHz, $2\gamma_2$=5 MHz and $2\gamma_3$=0.18 MHz, $\mu_{12}$=30 MHz and $\mu_{23}$=8 MHz. Here, $2\gamma_1$, $2\gamma_2$, $2\gamma_3$ are analogous to the losses of the cavity and the two polaritons in the three-level atomic medium, respectively; the coupling strength $\mu_{23}$ is in place of the Rabi frequency of the external "coupling" field $\Omega$ in the EIT terminology, and $\mu_{12}$ is similar to the cavity-atom coupling coefficient $g$ [40]. The predicted transmission spectrum looks nearly identical to the experimental results obtained in the intracavity EIT with cold atoms (Fig. 2 of Ref [40]): the upper/lower mode frequencies locate at $\pm 30.83$ MHz and the linewidth of the central mode is $2\gamma_c = 1.34$ MHz. As for the cavity #1 which is directly coupled to the waveguide, the central peak is narrowed more than 10 times. The linewidth narrowing is limited by $\gamma_3$, in analog to the ground-state decoherence rate in the three-level $\Lambda$-type atomic system [40]. Such three-peak intracavity EIT spectrum had also been observed in a system with Doppler-broadened three-level atoms inside an optical ring cavity [41].

In this TCPM platform, the central-peak narrowing can be further improved if a higher-Q microcavity (cavity #3) is employed and it can also be changed into a broadening case if a much lower-Q microcavity is used. Besides, by controlling the coupling strengths (and thus $\theta$), both the mode splitting and depths of the peaks in the spectrum can be easily modified. Furthermore, as shown in Fig.7 (a), the split side peaks and narrowed central peak can be separately detected if a probe tip is used to extract fields in cavity #2 and cavity #3, which will provide a convenient way for analyzing the spectral structures in this TCPM system.

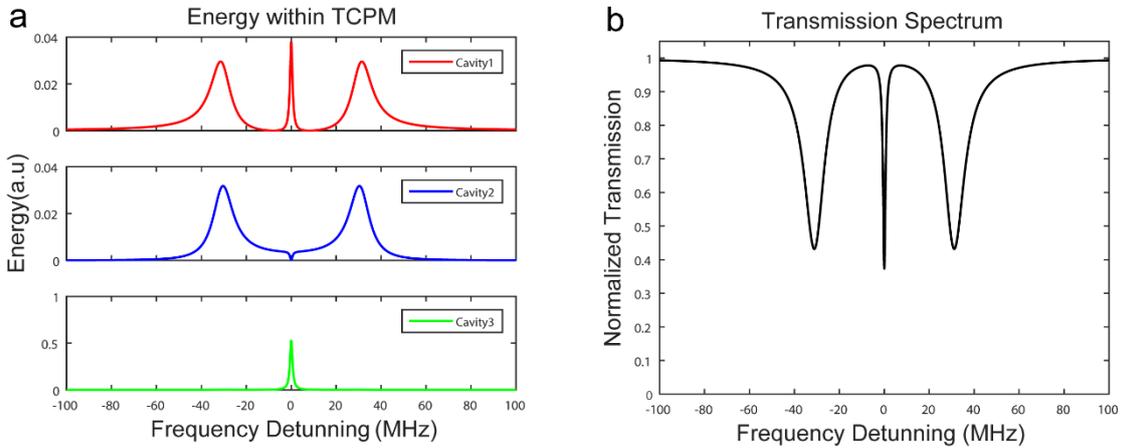

FIG.7 Calculated spectra on the analog to intracavity EIT in the TCPM platform. (a) The energy distributions in the three cavities, respectively. (b) The transmission spectrum through a waveguide side-coupled to cavity #1 showing the three-peak structure. Parameters are: $2\gamma_1 = 14$ MHz, $2\gamma_2 = 5$ MHz and $2\gamma_3 = 0.18$ MHz, $\mu_{12} = 30$ MHz and $\mu_{23} = 8$ MHz.

## VI. CONCLUSIONS

To evaluate the full field information (including phase and amplitude) in each cavity of a TCPM during forming and splitting of the supermodes, we have employed a Hamiltonian matrix approach [31] based on CMT. After examining the feasibility of the approach in the DCPM as a reference (all the calculations agree well with existing studies), we have analyzed the contributions of coherence and field distributions on mode splitting and linewidth evolutions, and thus pointed out the changes of fields around the exceptional point. This field-vector-based interpretation on the spectral properties of supermodes [31] opens a door to using such a simple Hamiltonian approach to study TCPM's optical properties and field information. Then, spectral properties of the TCPM, such as dark states, upper/lower mode splitting and linewidth evolutions, are calculated by the eigenvalues of the Hamiltonian matrix in cases of TCPM with three identical cavities, having an additional loss in the middle cavity, and finally having unequal losses in the cavities, respectively. Field properties in the cavities (including field amplitudes and phase differences) affected by the coupling strengths and additional loss in cavity #2 are fully discussed and a picture of interferences between the sub-DCPMs within the TCPM displays the interesting underlying physics of the TCPM structure. The central mode is shown to be promising in generating not only the robust "dark state" in the middle cavity, but also a phase shift of field after energy flowing between cavities #1 and #3 if the coupling strengths increase. Finally, we show an interesting analog between the TCPM and intracavity EIT (three-level atoms inside an optical cavity), in which the three-peak spectrum and substantially narrowed central peak occur in both systems with similar parameters. The peak narrowing can be controlled by changing coupling strengths and choosing proper cavity losses.

In summary, we have made a clear analysis on a chain-linked TCPM structure on its spectral properties and evolutions of field (energy) distributions. Interferences between sub-DCPMs in a TCPM were used to understand the energy distributions in the cavities and energy flowing between cavities in the structure with varying coupling strengths. Realization of a phase transition (or EP) is extended from DCPM to the TCPM. The demonstrated dark states and phase shifts in the system indicate that the abnormal central mode can find potential applications in fields like quantum information processing and remote control of energy in photonic devices. The tunable

TCPM is also a promising platform for spectral engineering and can be used to mimic strong interactions between multiple polaritons in solid-state and atomic systems. Topological studies by suitably designing modulations on coupling parameters and resonant frequencies of cavities around the EP might also be realized in such flexibly coupled multi-cavity systems.

## Acknowledgements

M.X. acknowledges the conversation with G.S. Agarwal on the issue of dark state in coupled cavities. Authors acknowledge supports from the National Basic Research Program of China (2012CB921804), the National Natural Science Foundation of China (nos. 61435007, 11574144 and 11321063), the Natural Science Foundation of Jiangsu Province (BK20150015), and the Fundamental Research Funds for the Central Universities.